# Impacts of Solar Intermittency on Future Photovoltaic Reliability

Jun Yin[1,2], Annalisa Molini[3], Amilcare Porporato[1,2]

[1]Department of Civil and Environmental Engineering, Princeton University, Princeton, New Jersey, 08544, USA.
[2]Princeton Environmental Institute, Princeton University, Princeton, New Jersey, 08544, USA.
[3]Masdar Institute, Khalifa University of Science and Technology, Abu Dhabi, UAE.
Corresponding author: Amilcare Porporato (aporpora@princeton.edu)

**Abstract**

As photovoltaic power is expanding rapidly worldwide, it is imperative to assess its promise under future climate scenarios. While a great deal of research has been devoted to trends of mean solar radiation, less attention has been paid to its intermittent character, a key challenge when compounded with uncertainties related to climate variability. Using both satellite data and climate model outputs, here we characterize solar radiation intermittency to assess future photovoltaic reliability. We find that the relation between the future power supply and long-term trends of mean solar radiation is highly nonlinear, thus making power reliability more sensitive to the fluctuations of mean solar radiation in regions where insolation is the highest. Our results highlight how reliability analysis must account simultaneously for the mean and intermittency of solar inputs when assessing the impacts of climate change on photovoltaics.

Increasing the use of solar energy is widely regarded as one of the most effective approaches to reduce $CO_2$ emissions, yet its intermittent nature imposes definite limitations to its reliability. While this problem may be partially solved by power storage, geographic dispersion, or load control[1–3], it already has significant impacts on the grid integration of renewable energy. For instance, photovoltaic power plants in Northwestern China (capacity of 43.87GW in 2019, 1/3 of China's total) were punished for providing intermittent energy to the Northwest Grid with fines of $0.2 billion yuan in 2017, $0.3 billion in 2018, and $0.27 billion for the first half of the year 2019, whereas coal-fired and hydropower plants were rewarded for their constant and even dispatchable sources of electricity[4–6]. Similarly, the example of Kauai island, Hawaii, a world pioneer in using renewable energy[7], currently relies on diesel generators on overcast days[8,9]. While the solar radiation varies across a range of timescale, here we focus on the daily level, which accounts for a significant portion of the penalty in the case of the Northwestern China[4,5] and related to the power reliability in Kauai, Hawaii[9].

The statistical properties of daily radiation are expected to change in future climates as a consequence of altered cloud and aerosol patterns[10–14]. Previous studies have focused mostly on the relative change of long-term mean radiation input[15–19]. While mean metrics are important, the portion of time with energy supply lower than the demand, termed loss-of-load probability (LOLP)[20], which is related to the reliability and the market values of power output, cannot be captured by mean values alone. As we will demonstrate here, in some cases, lower long-term mean solar radiation is associated with more reliable power outputs because of its lower intermittency.

To investigate the impacts of future climates on LOLP, we combine here satellite-derived data and climate model outputs. In particular, we focus on the impact of incident solar irradiance, one of the dominant factors controlling solar power generation[15,17,18].

**Characterizing solar energy intermittency**

We begin our investigation with an analysis of the clearness index, $K$, defined as the ratio between the near-surface global horizontal irradiance (GHI, including direct and diffuse irradiance) and the corresponding extraterrestrial horizontal irradiance (see Methods). This index accounts for the radiative and attenuation impacts from all factors such as clouds and aerosols and is often used in solar energy industry[21–24]. For example, we consider the case of Southeastern Romania, where climate change has shown strong regional impacts[25] and the case of Dubai, UAE, which is pursuing an ambitious plan to foster solar energy development in the region[26]. We use satellite data from Clouds and the Earth's Radiant Energy System (CERES), which have been used for solar power assessment[27,28]. Such decade-long records allow us to characterize the empirical distributions of daily $K$. As can be seen in Figure 1, the $K$ distributions for larger mean values (denoted as $\mu$ and also referred to as the mean clearness index) tend to have longer left tails, which are associated with the weaker solar radiation and lower power generation.

From the $K$ distribution, the LOLP of a solar power plant operating at daily basis (e.g., the Tesla's power plant at Kauai, Hawaii) can be estimated as the fraction of days with solar radiation lower than the demand value, $K_D$,

$$\text{LOLP} = \int_0^{K_D} f(K)dK , \qquad (1)$$

where $f(K)$ is the probability density function (pdf) of $K$ (see Methods). LOLP is therefore the cumulative density function (cdf) of $K$ at $K_D$. This type of metrics has long been used for designing a stand-alone (off-grid) photovoltaic power system[29–31] and is also a critical reference for evaluating a grid-connected system[20]. For example, the solar plant from Tesla is expected to provide 52 MWh of electricity every evening to the power grid in Kauai, Hawaii[7]. Tesla's design of 13MW solar array and 52 MWh effective battery storage result in an LOLP of 0.12, possibly maximizing the net profit while still satisfying the reliability requirement[9]. In a grid-connected system, LOLP is directly associated with the operating cost of the peaking plants (e.g., diesel generators in Kauai, Hawaii[8,9], hydropower stations in Northwest of China[32], gas turbines in the Great Plains, United States[33]) and thus linked to the market values of the solar energy. The constant demand $K_D$ in (1) is similar in spirit to the regulation from Northwest Grid of China, which was originally issued for coal plants considering their relatively constant power output but was recently adopted for solar and wind power plants. For all these reasons, a thorough characterization of the global solar power intermittency and its response to climate change using the LOLP is a fundamental starting point to assess the future reliability of photovoltaic.

Climate-change impacts on power reliability can be assessed by considering the change of LOLP during the lifespan of typical photovoltaic panels. Going back to the case of the Southern Romania, a solar plant designed under historical climate records of 2001-2009 is assumed to have a design LOLP, $LOLP_D$, of 0.3. Over the following nine years (2010-2018), the mean of $K$ increases in both January ($\Delta\mu$ =0.015) and July ($\Delta\mu$ =0.03) (indicated by the arrows in Figure 1 a and b). The corresponding values of LOLP drop from the design value of 0.3 to 0.27 in winter ($\Delta$LOLP = -0.03) and to 0.21 in summer ($\Delta$LOLP = -0.09), respectively (see the hatched and shaded areas in Figure 1 a and b). For the case in Dubai, $\Delta\mu$ and $\Delta$LOLP are found to be 0.025 and -0.09 in winter (Figure 1c) and negligible in summer (Figure 1d). This objectively quantifies not only the increase in mean surface solar radiation between these two periods (2001-2009 and 2010-2018), but also the increase in its reliability.

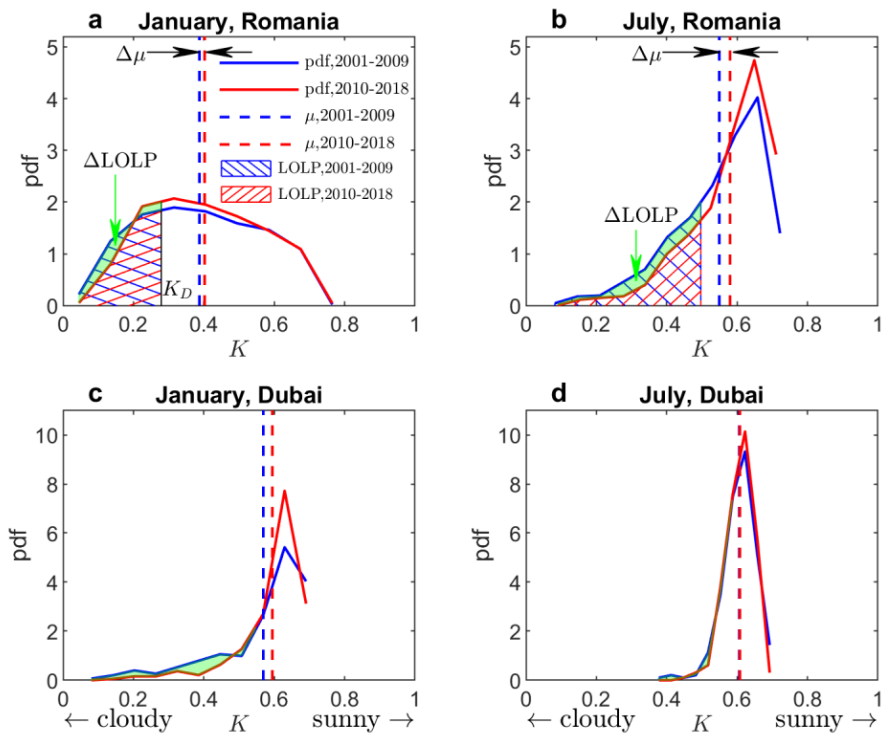

Figure 1. Examples of climate impacts on solar radiation and photovoltaic power reliability. The distribution of clearness index ($K$) derived from satellite data in (**a, c**) January and (**b, d**) July during 2001-2009 (blue lines) and during 2010-2018 (red lines) in (**a, b**) Southern Romania and (**c, d**) Dubai. The hatched areas indicate the probability when power generation does not meet the demand, the loss-of-load probability (LOLP, see text for details).

With this methodology, we now move to the future climate scenarios and use climate model outputs (see Supplementary Table 1) to calculate the changes of $\mu$ and LOLP between 2006-2015 and 2041-2050. As shown in Figure 2 (a and b) and in agreement with previous studies[15], the change of solar radiation is evident in some regions and show marked seasonal variations.

While the solar radiation in Europe is projected to decrease in January and increase in July, the radiation in Middle East decreases in both months. This redistribution of the Earth's energy and shifts in climate seasonality[34] have direct impacts on the solar power reliability as quantified by the corresponding variations of LOLP (see Figure 2 c and d). Although it is obvious that increasing solar radiation ($\Delta \mu > 0$) often leads to more reliable power output ($\Delta \text{LOLP} < 0$), this relationship is clearly nonlinear. For example, the slight decrease of solar radiation in Middle East and Northern Africa results in a significant increase of LOLP; increase of solar radiation in west of Amazon rainfall forest in July leads to sharp decrease of LOLP; strong variations in both radiation and power reliability are shown in Northern United States in January. In what follows, we will investigate this nonlinear relationship to quantitatively link our previous reports on mean solar radiation to one of our major concerns on power reliability.

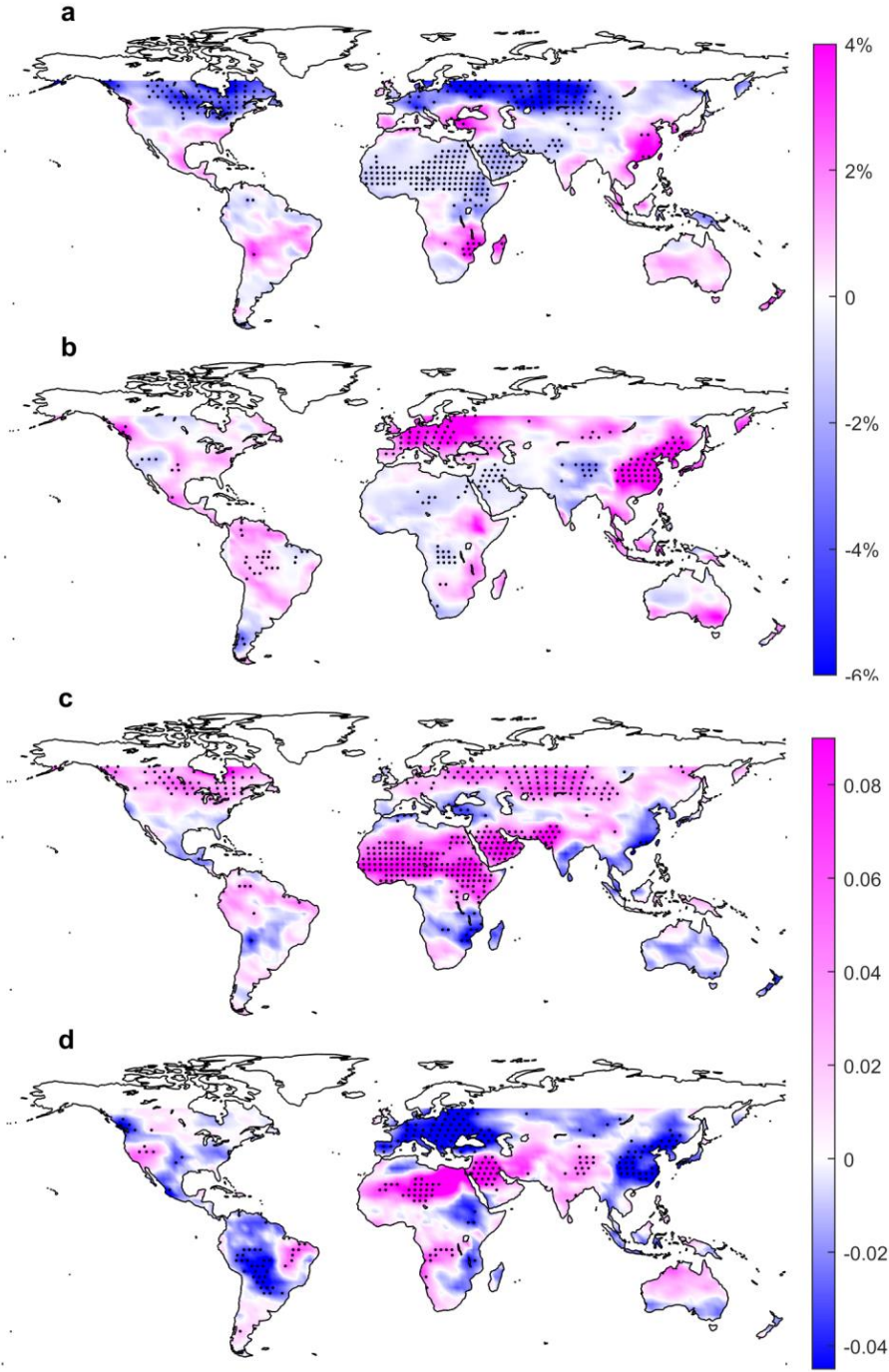

Figure 2. Variations of solar radiation and solar power reliability predicted from climate models. The color at each grid point represents the ensemble means of (**a, b**) $\Delta\mu/\mu$ and (**c, d**) $\Delta$LOLP between 2006-2015 and 2041-2050 in the month of (**a, c**) January and (**b, d**) July from 11 climate model outputs. The dots show the ensemble mean of the corresponding variables are statistically different than zero, suggesting consistent variations of solar radiation or reliability from most climate models (two sample *t*-test, 5% significance level). The LOLP during 2006-2015 (i.e., design LOLP) is set as 0.3; maps with other design LOLP show similar patterns (see Supplementary Fig. 1-2).

## Quantifying sensitivity of power reliability to climate change

The case studies in Figure 1 and geographical patterns in Figure 2 suggest that LOLP may be linked to the distribution of $K$, which in solar industry is often associated with the mean clearness index, $\mu$ [35,36]. To systematically assess this linkage, we consider in detail satellite data as well as climate model outputs under the historical climate conditions (see dark color curves in Figure 3a and Supplementary Fig. 3). As can be seen, $f(K)$ tends to be positively skewed in regions with smaller $\mu$ and negatively skewed in regions with larger $\mu$ (see Figure 3a); the standard deviation of $K$ (denoted as $\sigma$) tends first to slightly increase and then sharply decrease with $\mu$ (see Figure 3b). Moreover, these characteristics of $f(K)$ remains consistent in response to the changing climates for both short-term and long-term periods (see the light-color curves in Figure 3a and Supplementary Fig. 3). Overall, such empirical distributions even under changing climate conditions turn out to be well described by beta distributions (see Methods).

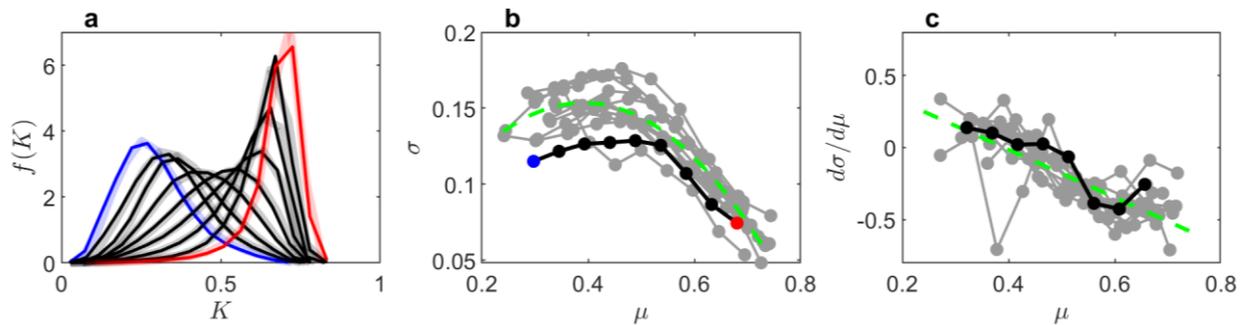

Figure 3. Dependence of distribution of clearness index, $K$, on its mean value, $\mu$. (**a**) Probability density functions (pdf) of daily clearness index ($K$) in different regions (e.g., the blue and red lines are from regions with monthly mean of $K$ ranges from 0.3 to 0.35 and from 0.65 to 0.70, respectively). The distributions of $K$ are from the satellite data in January during 2001-2009 (dark color) and during 2010-2018 (light color). (**b**) The relationship between mean ($\mu$) and standard deviation ($\sigma$) of daily $K$. The black, red, and blue dots correspond to the black, red, and blue lines in the panel (a). The grey dots/lines are from 11 climate model outputs (see Supplementary Table 1) of 'rcp45' experiment during 2006-2015; the dash green curve shows the best quadratic fit. (**c**) $d\sigma/d\mu$ calculated from the $\sigma \sim \mu$ relationships in panel (b) as the numerical (the gray or black dots/lines) or analytical (the dash green line) derivatives.

The fact that beta distributions provide a good fit to the solar radiation data allows us to link $\Delta\mu/\mu$ to $\Delta\text{LOLP}$ and thus in turn to obtain power-reliability information from previous reports on long-term mean solar radiation. Operationally, this can be accomplished by Taylor expanding Eq. (1) to first order as

$$\Delta \text{LOLP} \approx \underbrace{\mu \frac{\partial \text{LOLP}}{\partial \mu}}_{L_s} \left( \frac{\Delta \mu}{\mu} \times 100\% \right), \qquad (2)$$

where $L_s$ is the sensitivity of LOLP to $\mu$ and can be derived analytically for the beta distribution of $K$ (see Methods). In Eq. (2), the first term evaluates the climate impacts in terms of LOLP, whereas the term in the bracket assesses the future solar radiation in the conventional apporach[15–19]. The difference between the two, $\Delta \text{LOLP}$ and $\Delta \mu / \mu$, is clearly associated with the sensitivity parameter $L_s$, a nonlinear function of $\mu$ and $K_D$ (or design LOLP, see Eq. (11) in Methods). Particulary interesting is the fact that the absoulte values of $L_s$ are larger in sunny regions/seasons with larger $\mu$ (see Figure 4a). This may be accounted for by the fact that the small perturbation of $\mu$ in sunny regions tends to have larger change of variation of solar radiation (i.e., large absolute values of $d\sigma / d\mu$, see right side of Figure 3c), which is obviously associated with the intermittency of solar energy. Since these are also the regions of the World where the largest solar plants are expected to be deployed in the future, this fact should be considered with great attention in reliability analysis.

The previous analytical results are corroborated by climate model outputs. Figures 3d shows $\Delta \mu / \mu$ and $\Delta \text{LOLP}$ between 2006-2015 to 2041-2050 for given values of $\mu$ and design LOLP. The slopes of these two quantities are reported in Figure 4b, showing similar patterns as their analytical counterparts (Figure 4a).

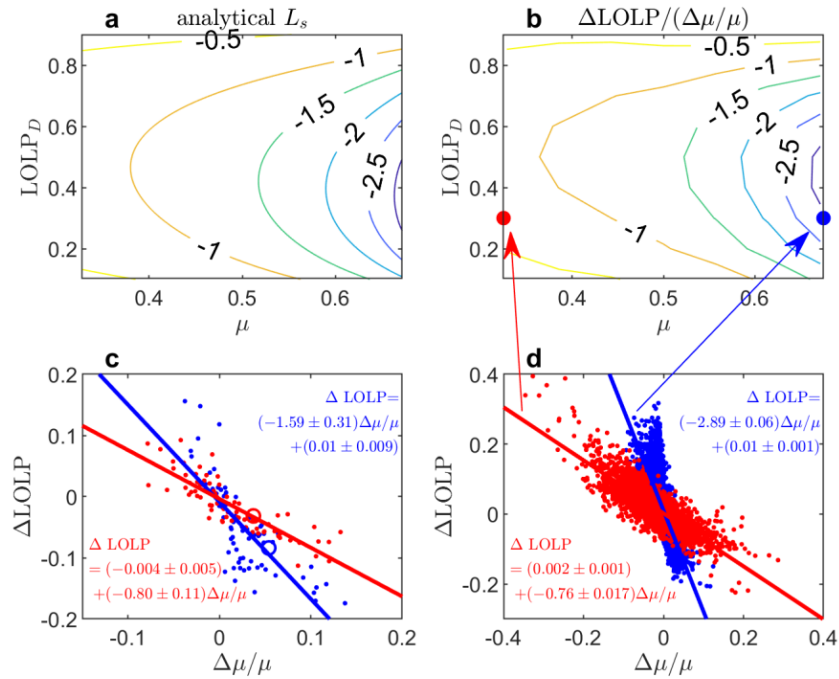

Figure 4. Sensitivity of loss-of-load probability ($L_s$). Contour plots of $L_s$ is calculated (**a**) analytically from Eq. (2) and (**b**) numerically from climate model outputs. The red and blue dots in (b) are corresponding to the examples in (**d**), which compares the change of LOLP and the change of $\mu$ from 2006-2015 to 2041-2050 in January with design LOLP of 0.3 in regions

where $0.3 < \mu < 0.35$ (red dots) and $0.65 < \mu < 0.7$ (blue dots) as projected by climate models. The red and blue lines are the corresponding best fit lines and their slopes (i.e., $\Delta \text{LOLP} / (\Delta \mu / \mu)$) numerically represent $L_s$. (**c**) As in (d) but only in Southeast Europe in January (red dots) and July (blue dots). The red and blue circles correspond to the example of Southern Romania in Figure 1.

With the obtained nonlinear function of $L_s$ (see Eq. (10) in Methods), one can readily infer the power reliability. For example, Figure 5 shows that $L_s$ is approximately -0.8 in January and -1.6 in July in Southern Romania for a design LOLP of 0.3. The mean solar radiation in this region is projected to vary around -15~0% in winter and around -5~5% in summer toward the end of the century[18]. Multiplying these variations by $L_s$, one can find the impacts of these variations on LOLP (i.e., 0~12% in winter and -8~8% in summer). While the winter season has larger variations in solar radiation, it also has small absolute value of $L_s$ so that the impacts on future power reliability in winter are reduced. This analysis is also consistent with the results from climate model outputs as shown in Figure 4c, which suggests larger spread of $\Delta$ LOLP but slightly smaller change of $\Delta \mu / \mu$ in summer.

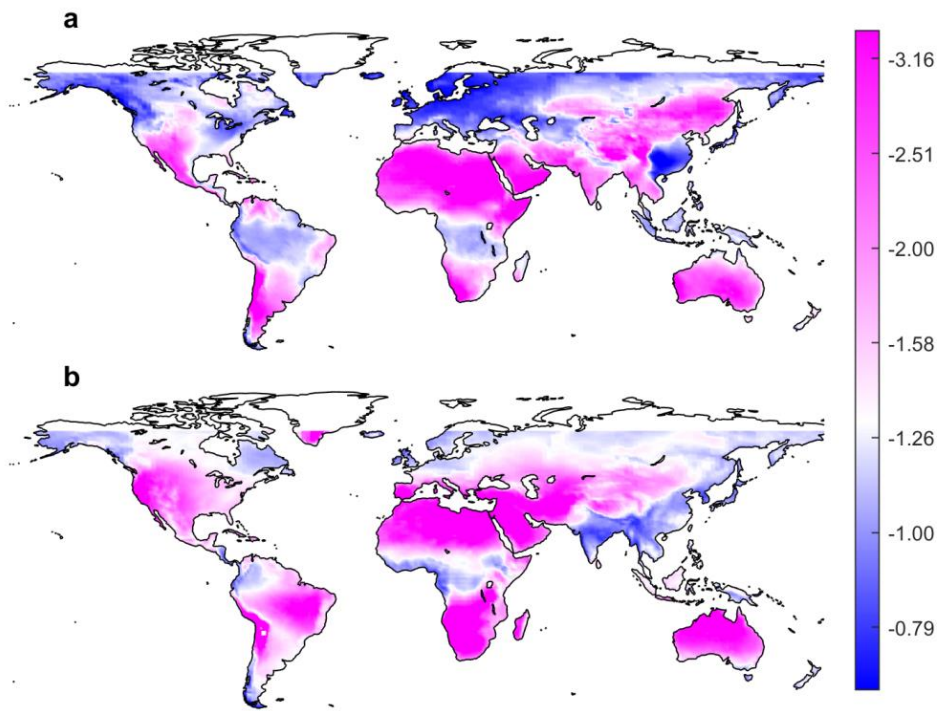

Figure 5. Global maps of LOLP sensitivity ($L_s$). This sensitivity in (**a**) January and (**b**) July is obtained from analytical solutions with design LOLP of 0.3 and solar radiation climatology from CERES. Maps with different design LOLP values show similar spatial patterns (see Supplementary Fig. 4-5).

**Discussion**

The global maps of $L_s$ in Figure 5, similarly to the cloud climatology patterns previously reported[37], show that power reliability is less sensitive to mean solar radiation in wet regions (e.g., Europe, Southeast China, Southeast United States) and more sensitive in dry regions (Middle East and Northern Africa). This is consistent with the observed $\sigma \sim \mu$ relationship in Figure 3 b, where these slopes are steeper in the sunny regions. Meanwhile, the wet regions are predicted to have relatively more solar radiation in the future climate scenarios[15,17]. The multiplication of small negative $L_s$ with positive $\Delta\mu$ yields small but negative $\Delta \text{LOLP}$, suggesting slightly higher power reliability. On the other hand, the dry regions are predicted to have slightly less solar radiation but could yield much lower power reliability due to the strong LOLP sensitivity. Therefore, this sensitivity factor reveals a strong nonlinear relationship between mean solar and power intermittency (Figure 2) which had not emphasized previously.

In summary, our results have shown how the impacts of this radiation change on power reliability could be significant due to the large absolute values of LOLP sensitivity. This point towards a tradeoff between the mean solar radiation that quantifies the total potential solar power and the power reliability, which being related to intermittency remains a major concern in the absence of large power storage options. This contrasting behavior between solar power availability and reliability requires special attention in assessments of future solar energy scenarios.

**Reference**

1. Fripp, M. Switch: A Planning Tool for Power Systems with Large Shares of Intermittent Renewable Energy. *Environ. Sci. Technol.* **46**, 6371–6378 (2012).
2. He, G. *et al.* SWITCH-China: A Systems Approach to Decarbonizing China's Power System. *Environ. Sci. Technol.* **50**, 5467–5473 (2016).
3. Perez, R. *et al.* Achieving very high PV penetration – The need for an effective electricity remuneration framework and a central role for grid operators. *Energy Policy* **96**, 27–35 (2016).
4. Northwest China Energy Regulatory. Regulation for grid-connected power plants in Northwest regions. http://xbj.nea.gov.cn/website/Aastatic/news-196276.html (2018).
5. Ying, L. Northwest photovolatic power plants received $0.27 billion ticket for the first half year of 2019. *CNGOLD* https://energy.cngold.org/c/2019-09-19/c6584966.html (2019).
6. Guangfumen. Penalty doubled: photovoltaic and wind power plants in Qinghai receive 17 million tickets in January. *Sohu* http://www.sohu.com/a/314591748_609294.
7. KIUC. Renewables | Kauai Island Utility Cooperative. http://website.kiuc.coop/renewables (2019).
8. Golson, J. Tesla built a huge solar energy plant on the island of Kauai. *The Verge* https://www.theverge.com/2017/3/8/14854858/tesla-solar-hawaii-kauai-kiuc-powerpack-battery-generator (2017).
9. Klippenstein, M. Tesla's solar and battery project in Hawaii: we do the math. *Green Car Reports* https://www.greencarreports.com/news/1112800_teslas-solar-and-battery-project-in-hawaii-we-do-the-math (2017).

10. Dessler, A. E. A Determination of the Cloud Feedback from Climate Variations over the Past Decade. *Science* **330**, 1523–1527 (2010).
11. Boucher, O. *et al.* Clouds and Aerosols. in *Climate Change 2013: The Physical Science Basis. Contribution of Working Group I to the Fifth Assessment Report of the Intergovernmental Panel on Climate Change* (eds. Stocker, T. F. et al.) 571–658 (Cambridge University Press, 2013).
12. Bloomfield, H. C., Brayshaw, D. J., Shaffrey, L. C., Coker, P. J. & Thornton, H. E. Quantifying the increasing sensitivity of power systems to climate variability. *Environ. Res. Lett.* **11**, 124025 (2016).
13. Yin, J. & Porporato, A. Diurnal cloud cycle biases in climate models. *Nat. Commun.* **8**, 2269 (2017).
14. Yin, J. & Porporato, A. Radiative effects of daily cloud cycle: general methodology and application to cloud fraction. *ArXiv Phys.* (2018).
15. Crook, J. A., Jones, L. A., Forster, P. M. & Crook, R. Climate change impacts on future photovoltaic and concentrated solar power energy output. *Energy Environ. Sci.* **4**, 3101–3109 (2011).
16. Bartos, M. D. & Chester, M. V. Impacts of climate change on electric power supply in the Western United States. *Nat. Clim. Change* **5**, 748–752 (2015).
17. Wild, M., Folini, D., Henschel, F., Fischer, N. & Müller, B. Projections of long-term changes in solar radiation based on CMIP5 climate models and their influence on energy yields of photovoltaic systems. *Sol. Energy* **116**, 12–24 (2015).
18. Jerez, S. *et al.* The impact of climate change on photovoltaic power generation in Europe. *Nat. Commun.* **6**, 10014 (2015).
19. Bazyomo, S. D. Y. B., Agnidé Lawin, E., Coulibaly, O. & Ouedraogo, A. Forecasted Changes in West Africa Photovoltaic Energy Output by 2045. *Climate* **4**, 53 (2016).
20. NERC. *Probabilistic Adequacy and Measures*. (2018).
21. Aguiar, R. J., Collares-Pereira, M. & Conde, J. P. Simple procedure for generating sequences of daily radiation values using a library of Markov transition matrices. *Sol. Energy* **40**, 269–279 (1988).
22. Lambert, T., Gilman, P. & Lilienthal, P. Micropower system modeling with HOMER. *Integr. Altern. Sources Energy* 379–418 (2005).
23. Yadav, A. K. & Chandel, S. S. Tilt angle optimization to maximize incident solar radiation: A review. *Renew. Sustain. Energy Rev.* **23**, 503–513 (2013).
24. Bett, P. E. & Thornton, H. E. The climatological relationships between wind and solar energy supply in Britain. *Renew. Energy* **87**, 96–110 (2016).
25. Kayser-Bril, N. Europe is getting warmer, and it's not looking like it's going to cool down anytime soon. *European Data Journalism Network* https://www.europeandatajournalism.eu/eng/News/Data-news/Europe-is-getting-warmer-and-it-s-not-looking-like-it-s-going-to-cool-down-anytime-soon (2018).
26. IRENA. *Renewable Energy Market Analysis: GCC 2019*. (2019).
27. Chandler, W. S., Whitlock, C. H. & Stackhouse, P. W. NASA climatological data for renewable energy assessment. *J. Sol. Energy Eng.* **126**, 945–949 (2004).
28. Li, X., Wagner, F., Peng, W., Yang, J. & Mauzerall, D. L. Reduction of solar photovoltaic resources due to air pollution in China. *Proc. Natl. Acad. Sci.* **114**, 11867–11872 (2017).
29. Barra, L., Catalanotti, S., Fontana, F. & Lavorante, F. An analytical method to determine the optimal size of a photovoltaic plant. *Sol. Energy* **33**, 509–514 (1984).

30. Chapman, R. N. *Sizing handbook for stand-alone photovoltaic/storage systems*. (Sandia National Laboratories, 1987).
31. Egido, M. & Lorenzo, E. The sizing of stand alone PV-system: A review and a proposed new method. *Sol. Energy Mater. Sol. Cells* **26**, 51–69 (1992).
32. Ming, B. *et al.* Optimizing utility-scale photovoltaic power generation for integration into a hydropower reservoir by incorporating long- and short-term operational decisions. *Appl. Energy* **204**, 432–445 (2017).
33. DeCarolis, J. F. & Keith, D. W. The economics of large-scale wind power in a carbon constrained world. *Energy Policy* **34**, 395–410 (2006).
34. Feng, X., Porporato, A. & Rodriguez-Iturbe, I. Changes in rainfall seasonality in the tropics. *Nat. Clim. Change* **3**, 1–5 (2013).
35. Liu, B. Y. H. & Jordan, R. C. The interrelationship and characteristic distribution of direct, diffuse and total solar radiation. *Sol. Energy* **4**, 1–19 (1960).
36. Klein, S. A. & Beckman, W. A. Loss-of-load probabilities for stand-alone photovoltaic systems. *Sol. Energy* **39**, 499–512 (1987).
37. Rossow, W. B. & Schiffer, R. A. ISCCP Cloud Data Products. *Bull. Am. Meteorol. Soc.* **72**, 2–20 (1991).
38. Gunerhan, H. & Hepbasli, A. Determination of the optimum tilt angle of solar collectors for building applications. *Build. Environ.* **42**, 779–783 (2007).
39. Vaillon, R., Dupré, O., Cal, R. B. & Calaf, M. Pathways for mitigating thermal losses in solar photovoltaics. *Sci. Rep.* **8**, 13163 (2018).
40. Bendt, P., Collares-Pereira, M. & Rabl, A. The frequency distribution of daily insolation values. *Sol. Energy* **27**, 1–5 (1981).
41. Hollands, K. G. T. & Huget, R. G. A probability density function for the clearness index, with applications. *Sol. Energy* **30**, 195–209 (1983).
42. Davison, A. C. *Statistical models*. vol. 11 (Cambridge University Press, 2003).

**Acknowledgment**

We would like to thank Professors Robert Socolow and Tiejian Li for their constructive comments on this work. J.Y. and A.P. acknowledge support from the USDA Agricultural Research Service cooperative agreement 58-6408-3-027; and National Science Foundation (NSF) grants EAR-1331846, EAR-1316258, FESD EAR-1338694 and the Carbon Mitigation Initiative at Princeton University. A.M. acknowledges support from the Khalifa University Competitive Internal Research Award, CIRA-2017-102.

**Methods**

**Clearness Index ($K$)**

The daily clearness index, $K$, is defined as

$$K = \frac{\int_0^T \text{GHI}(t)dt}{\int_0^T \text{EHI}(t)dt}, \tag{3}$$

where $T$ is the length of one day, GHI is the near-surface global horizontal irradiance, and EHI the extraterrestrial horizontal irradiance. Daily GHI are obtained from Clouds and the Earth's Radiant Energy System (CERES) SYN1deg during 2001-2018 and from 11 climate model outputs (ACCESS1.3, BCC-CSM1.1m, CanESM2, CCSM4, CMCC-CMS, CSIRO-Mk3.6.0, EC-EARTH, GFDL-CM3, INM-CM4, IPSL-CM5A, and MPI-ESM) in 'rcp45' experiment during 2006-2015 and 2041-2050. All these data have been used to obtain the empirical distributions of $K$ for calculating the loss-of-load probability as explained next.

**Loss-Of-Load Probability (LOLP)**

The photovoltaic power output is related to the incident solar radiation and other factors controlling the solar cell efficiency[15]. In each month, the Sun's declination angle has small variations; the daily incident solar radiation on a fixed or tracking array can be approximated as a monotonical function of daily clearness index[38]. Factors such as dust and tree shading on solar arrays could have notable impacts on power generation but can be controlled by regular maintanence. The solar cell efficiency factors such as air temperature and wind speed usually have only a secondary impacts and can be controlled by multiple mitigation strategies[39]. In regard to climate change impacts, the incident solar radiation has been identified as the dominant factor for photovoltaic power generation (e.g., see the spatial patterns in Fig.1 of Crook et al., 2011 and Fig.1 of Jerez et al., 2015). For this reason, we model the power output as a monotonic function of clearness index, say $p = g(K)$. This function can be used to estimate the loss-of-load probability (LOLP). For a power system with daily storage capacity, LOLP can be defined as the fraction of days when daily energy supply ($p$) is lower than the daily demand ($p_D$). We obtain LOLP as the derived distribution of $K$,

$$\text{LOLP} = F(p_D = g(K_D)) = F(K_D) = \int_0^{K_D} f(K)dK, \tag{4}$$

where $K_D$ is the specific value of $K$ that is just enough to generate the demanding energy $p_D$, $f(\cdot)$ and $F(\cdot)$ are the probability and cumulative density function of $K$. These functions are estimated from multi-year historical climate records, and thus the corresponding LOLP already captures the interannual variability of daily power generation. Such estimates are referred to as design LOLP, $\text{LOLP}_D$. For the lifespan of a typical photovoltaic panel (20-30 years), one can then quantify the climate impacts on power reliability as the change of LOLP from its design value.

**LOLP Sensitivity ($L_s$)**

The distributions of $K$ enters the LOLP expression in Eq. (4). As presented in Figure 3a, the distribution of $K$ tends to be positively skewed for smaller mean value of $K$ (denoted as $\mu$) and negatively skewed for larger $\mu$. These types of distributions in the United States have been described by several empirical expressions[40,41]. To better fit the satellite observations across the world, here we model the distribution of $K$ as the beta distribution

$$f_b(K; \beta_1, \beta_2) = \frac{\Gamma(\beta_1 + \beta_2)}{\Gamma(\beta_1)\Gamma(\beta_2)} K^{\beta_1 - 1}(1 - K)^{\beta_2 - 1}, \tag{5}$$

where $\beta_1$ and $\beta_2$ are the shape parameters. These shape parameters can be expressed by the mean ($\mu$) and standard deviation ($\sigma$) of the distribution[42],

$$\beta_1 = \frac{\mu(\mu - \mu^2 - \sigma^2)}{\sigma^2}, \tag{6}$$

and

$$\beta_2 = \frac{(1 - \mu)(\mu - \mu^2 - \sigma^2)}{\sigma^2}. \tag{7}$$

As described in Figure 3b, the standard deviation may be modeled as a function of mean (e.g., $\sigma = -0.83\mu^2 + 0.65\mu + 0.03$, the best quadratic fit) so that the distribution of $K$ can be written as

$$f_b(K; \beta_1, \beta_2) = f_b(K; \beta_1(\mu, \sigma(\mu)), \beta_2(\mu, \sigma(\mu))). \tag{8}$$

Substituting (8) into (4) and performing a Taylor expansion to first order yields

$$\Delta \text{LOLP} \approx \underbrace{\mu \frac{\partial \text{LOLP}}{\partial \mu}}_{L_s} \left( \frac{\Delta \mu}{\mu} \times 100\% \right), \tag{9}$$

where

$$L_s = \mu \left( \frac{\partial \sigma}{\partial \mu} \frac{\partial \beta_1}{\partial \sigma} + \frac{\partial \beta_1}{\partial \mu} \right) \frac{\partial F}{\partial \beta_1}\bigg|_{K=K_D} + \mu \left( \frac{\partial \sigma}{\partial \mu} \frac{\partial \beta_2}{\partial \sigma} + \frac{\partial \beta_2}{\partial \mu} \right) \frac{\partial F}{\partial \beta_2}\bigg|_{K=K_D}, \tag{10}$$

where $F_b(\cdot)$ is the cumulative beta distribution and $K_D$ is equivalent to design LOLP,

$$\text{LOLP}_D = F_b(K_D). \tag{11}$$

The corresponding analytical solutions of $L_s$ (Figure 4a) are very similar to its counterpart calculated numerically as $\Delta \text{LOLP} / (\Delta \mu / \mu)$ (Figure 4b).

**Data Availability**

Supplementary Information for

# Impacts of Solar Intermittency on Future Photovoltaic Reliability

Jun Yin, Annalisa Molini, Amilcare Porporato

The following figures and table provide complementary information regarding reliability of solar power and its response to climate change.

- Supplementary Fig. 1-2 show the climate impacts on the power reliability.
- Supplementary Fig. 3 shows the distribution of clearness index from different climate models.
- Supplementary Fig. 4-5 shows the sensitivity of climate impacts on power reliability.
- Supplementary Table 1 lists all the model names in this study.

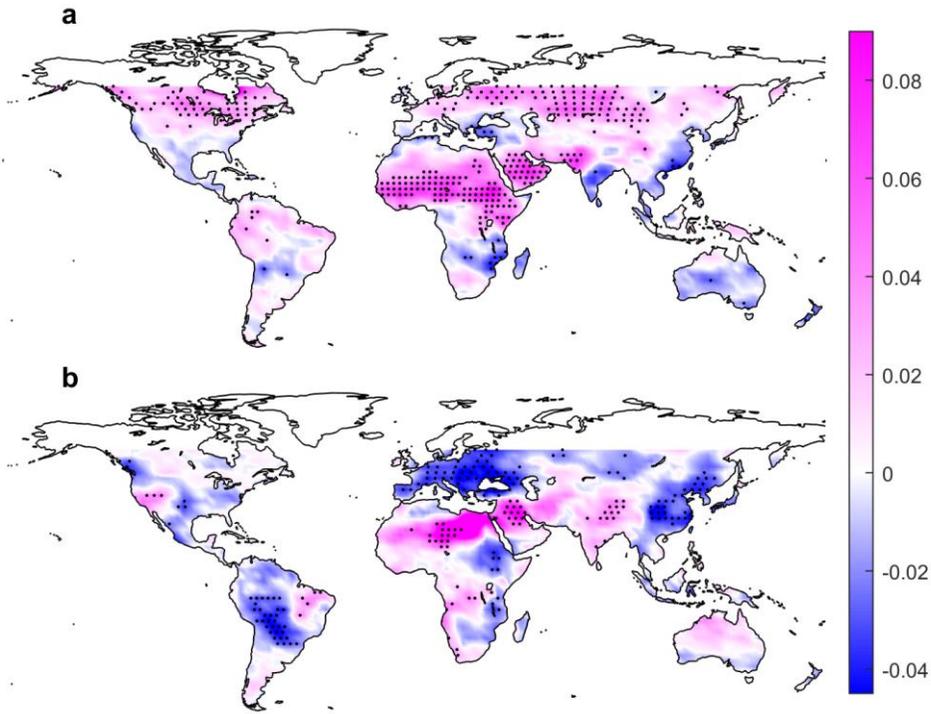

Supplementary Fig. 1. Ensemble means of ΔLOLP as in Figure 2 c and d in the main text but for design LOLP of 0.2 in (a) January and (d) July.

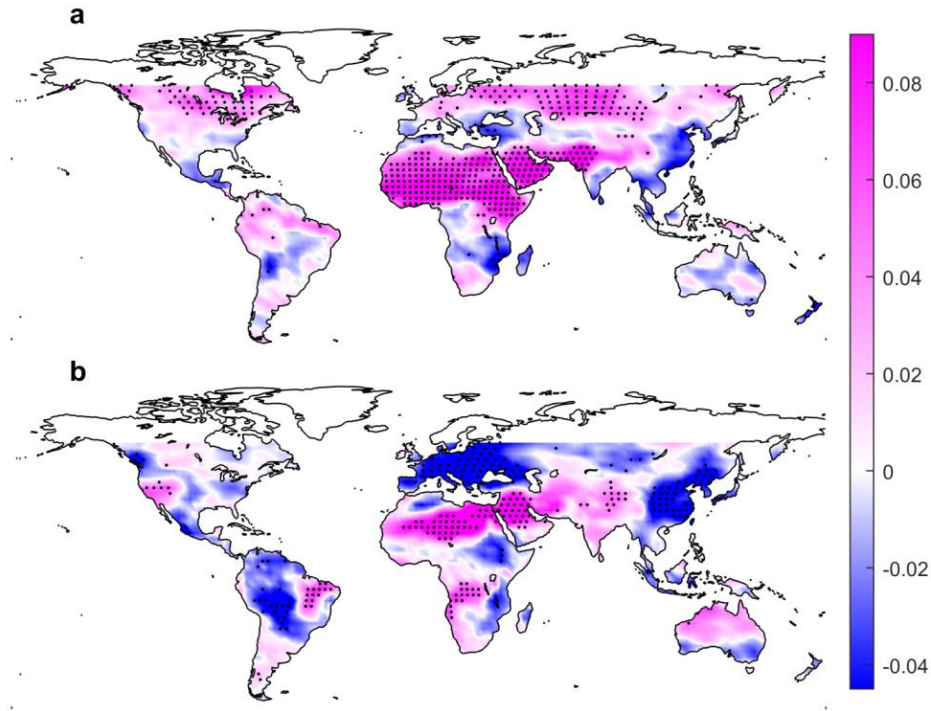

Supplementary Fig. 2. Ensemble means of ΔLOLP as in Figure 2 c and d in the main text but for design LOLP of 0.4 in (a) January and (d) July.

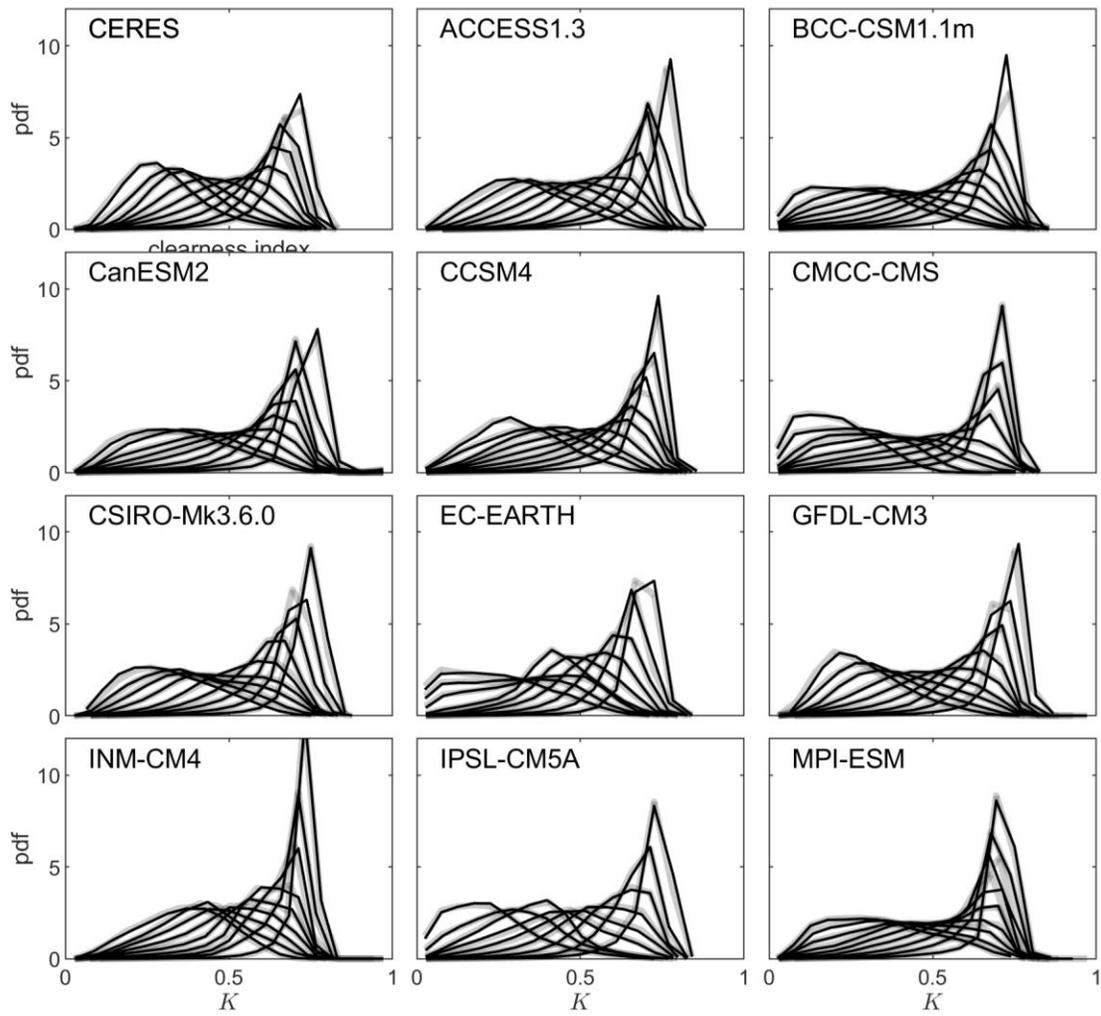

Supplementary Fig. 3. The first panel is the same as Figure 3a in the main text and the rest are referred to the distributions of *K* from 11 climate model outputs during 2006-2015 (dark color) and 2041-2050 (light color) in 'rcp45' experiment.

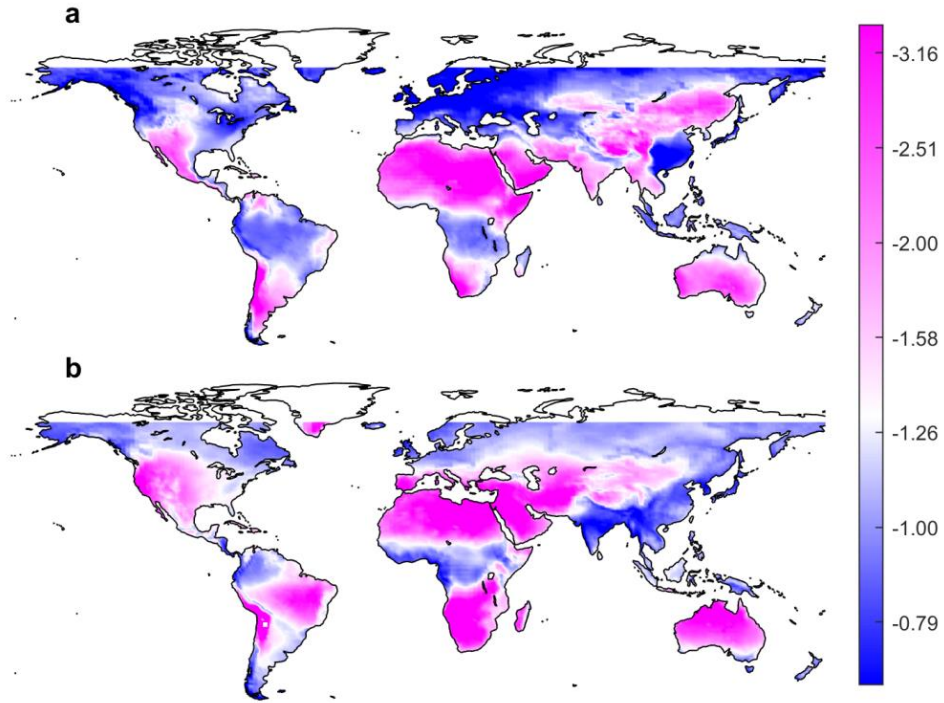

Supplementary Fig. 4. As in Figure 5 in the main text but for design LOLP of 0.2.

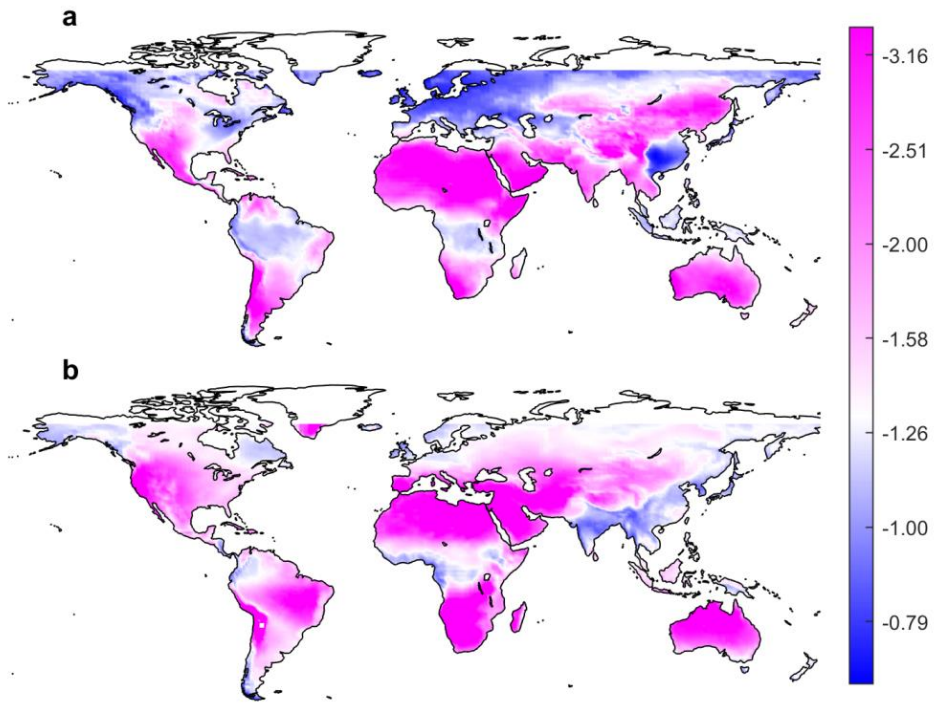

Supplementary Fig. 5. As in Figure 5 in the main text but for design LOLP of 0.4.

Supplementary Table 1. Climate Models used for assessing the intermittency and reliability of photovoltaic power in this study.

| Acronyms | Model Institutions |
| --- | --- |
| ACCESS1.3 | Commonwealth Scientific and Industrial Research Organization, Australia |
| BCC-CSM1.1(m) | Beijing Climate Center, China |
| CanESM2 | Canadian Centre for Climate Modelling and Analysis, Canada |
| CCSM4 | National Center for Atmospheric Research, USA |
| CMCC-CMS | Euro-Mediterranean Center on Climate Change, Italy |
| CSIRO-Mk3.6.0 | Commonwealth Scientific and Industrial Research Organization, Australia |
| EC-EARTH | EC-Earth consortium, Europe |
| GFDL-CM3 | NOAA Geophysical Fluid Dynamics Laboratory, USA |
| INM-CM4 | Institute for Numerical Mathematics, Russia |
| IPSL-CM5A | Institute Pierre Simon Laplace, France |
| MPI-ESM-MR | Max Planck Institute for Meteorology (MPI-M), Germany |